\documentclass[11pt,onecolumn]{article}
\usepackage[a4paper, total={6.5in, 8.5in}]{geometry}
\usepackage{graphicx}
\usepackage{unicode-math}
\usepackage{nicefrac}
\usepackage{xcolor}
\usepackage{url}
\usepackage{hyperref}
\usepackage{cite}
\usepackage{multicol}
\usepackage{multirow}
\usepackage{float}
\usepackage[width=0.8\linewidth,font={small}]{caption}
\usepackage[width=0.8\linewidth,font={small}]{subcaption}
\usepackage{setspace}
\usepackage{abstract}
\DeclareCaptionLabelFormat{andtable}{#1~#2  \&  \tablename~\thetable}

\usepackage{authblk}

\begin{document}
\title{Local Structure of Epitaxial Single Crystal UO$_{2+x}$ Thin Films}

\author[1,*]{Jarrod C. Lewis}
\author[2,3]{Steven D. Conradson}
\author[1]{Jacek Wasik}
\author[1]{Lottie M. Harding}
\author[1]{Rebecca Nicholls}
\author[4]{Jude Laverock}
\author[1]{Chris Bell}
\author[1]{Ross Springell}
\affil[1]{School of Physics, University of Bristol, Bristol, BS8 1TL, UK.}
\affil[2]{Department of Complex Matter, Josef Stefan Institute, 100 Ljubljana, Slovenia.}
\affil[3]{School of Chemistry, University of Bristol, Bristol, BS8 1TS, UK.}
\affil[4]{Department of Chemistry, Washington State University, Pullman, WA 90164, USA.}
\affil[*]{Current address: Department of Materials, University of Oxford, Oxford, OX1 3PH, UK.}
\date{}
\maketitle

\maketitle
\begin{abstract}

The influence of oxygen stoichiometry on the uranium local environment is explored in epitaxial single crystal uranium oxide thin films grown by DC magnetron sputtering. Through post-growth annealing, the stoichiometry of as-grown UO$_{2}$ films are tuned over an approximate stoichiometry range of $0.07 \leq x \leq 0.20$, estimated with X-ray photoelectron spectroscopy measurements of the U$-4f$ and O$-1s$ peaks. The local structure of the thin films are then probed using extended X-ray absorption fine structure measurements at the U $L_{3}$ absorption edge. We observe both the evolution of the U local environment of as a function of oxidation in UO$_{2+x}$, and that the near stoichiometric UO$_{2}$ film replicates the local structure of bulk UO$_{2}$ material standards well. The series of stoichiometrically varied samples highlights the non-trivial transitional behaviour of the UO$_{2+x}$ oxygen sublattice with increasing oxygen content in this stoichiometric regime, while also demonstrating the efficacy of this thin film synthesis route for actinide studies beyond their established use as idealised surfaces, which could be readily adapted for further stoichiometrically tailored material studies and UO$_{2+x}$ device fabrication. 

\end{abstract}

\section{Introduction}

The binary uranium oxide phase diagram is rich in both structural diversity and chemical complexity \cite{McEachernandTaylor1998,Idriss2010}, with continuing developments even after decades of study \cite{Baichi2006,Prieur2021,Wasik2024,Silva2024}. The evolution of material properties with the variation of oxide stoichiometry $x$ for a parent compound of UO$_{2+x}$ is a substantial body of work. Recently there are a number of reports noting the possible utility of these oxygen content-dependent behaviours for the design of functional materials and devices \cite{Meek2008,Kruschwitz2014,Young2016,Liu2019,Rickert2021,Cheng2022,Xie2023}. Of particular interest is the stoichiometric range of $0<x<0.67$, which is bound on either end by the single-valence U(IV) face-centred cubic UO$_{2}$ phase ($x=0$, space group $Fm\bar{3}m$), and the mixed-valence U(V)/U(VI) orthorhombic U$_{3}$O$_{8}$ phase ($x=0.67$, space group $C2mm$) \cite{Leinders2017,Hoekstra1961,Loopstra1970}. The magnetic, optical and electronic structure properties of both UO$_{2}$ \cite{Caciuffo2011,An2011,Conradson2013,Bao2013,Gilbertson2014,Tereshina2015,Conradson2016,Ine2024,Thomas2024} and U$_{3}$O$_{8}$ \cite{Saniz2023,LawrenceBright2023,Miskowiec2024} attract continued research interest from a fundamental material science perspective. For UO$_{2+x}$ between these two stoichiometric compounds the picture is less clear. New insights are frequently reported, related to both the crystal structures adopted by the material, and the properties associated with the variation the U local environment with changing oxygen content  \cite{Conradson2004,ElJamal2022,Elorrieta2022,Leinders2024,Lewis2024}. 

\par 

However, it can be challenging to relate observations made using different uranium oxide sample morphologies, such as sintered powders and bulk single crystals. While powder based synthesis routes with rich microstructural variation continue to attract attention due to their intrinsic alignment to studies of nuclear fuel materials, work with solid crystalline samples is much less prolific, in part motivated by how challenging both growing and handling bulk single crystals of actinide materials can be. As such, there is continued interest in exploring how the rich material physics reported in sintered material systems with varying oxygen content manifests within uranium oxide materials with extended crystalline ordering, particularly in the UO$_{2+x}$ stoichiometry regime. 

\par

To study solid UO$_{2+x}$, we synthesise epitaxial thin film single crystals, providing highly ordered systems in which to study small structural variations without the requirement of uniform bulk crystals. Following initial characterisation of the as-grown crystal structure with X-ray diffraction (XRD), and valence state characterisation using X-ray photoelectron spectroscopy (XPS), we studied the volume average local structure of a series of UO$_{2+x}$ thin films. This made use of X-ray absorption fine structure (XAFS) measurements at the U $L_{3}$ absorption edge ($E=17.166\:\mathrm{keV}$), and analysis of the extended X-ray absorption fine structure (EXAFS) oscillations within the X-ray fluorescence signal. The inter-atomic distances within the sample are characterised both as a function of the annealing treatment and sample temperature during the measurements. These are then fitted with a stoichiometric UO$_{2}$ single-scattering model to probe the variation in the local structure of the thin film from that of stoichiometric UO$_{2}$.

\par 

We have also made similar measurements for bulk powder UO$_{2}$ to explore the ability of these thin films to replicate the bulk local structure. Through this combination of characterisation techniques, it is demonstrated that this epitaxial thin film synthesis route enables the replication of the complex anionic behaviour observed in the bulk material, and that this is further tuneable by post growth annealing.

\section{Experimental Details}

The samples were grown by reactive DC magnetron sputtering from a depleted uranium target in an ultra-high vacuum deposition chamber with a base pressure of the order $10^{−8}$ Pa. This dedicated actinide deposition system is part of the FaRMS facility at the University of Bristol \cite{farms, Springell2022}. All samples used calcium fluoride $\left(\mathrm{CaF}_{2}\right)$ single crystals as substrates, sourced from MTI Corp\textsuperscript{\textregistered}, with dimensions of $10$ mm $\times$ $10$ mm $\times$ $0.5$ mm, and polished to optical grade with a root-mean-square surface roughness of $\leq5$ \AA. The high degree of epitaxial matching between CaF$_{2}$ and stoichiometric UO$_{2}$ facilitates epitaxial growth with minimal strain or crystallographic dislocations \cite{RennieThesis}. UO$_{2}$ layers were deposited on $[111]$ or $[100]$ oriented CaF$_{2}$ substrates, producing single crystal UO$_{2}$ thin films with corresponding orientations. 

\par 

During UO$_{2}$ growth at a sputtering target power of $50$ W, partial pressures of Ar and O$_{2}$ were maintained at $7.7\times10^{−1}$ and $4\times10^{−3}$ Pa respectively, and the substrate temperature was fixed at $500 ^\circ$C. The substrate was annealed at this temperature for $1$ hour prior to deposition to desorb any surface contaminants and to thermalise the substrate. To provide a sufficient signal amplitude for fluorescence XAFS measurements, layers were grown to a nominal thickness of $500$ nm, as calibrated using X-ray reflectivity to a rate of $\approx 3.8$ nm s$^{-1}$. 
Samples of nominal initial UO$_{2}$ ($x=0$) stoichiometry were annealed to alter the oxygen content of the thin film using one of two set ups. For characterisation in XRD and XAFS experiments, samples were annealed under various conditions detailed in Table \ref{tab:Table1} on a heated sample stage inside of an Anton Paar HTK-1200N high temperature oven. 

\begin{table}[h!]
    \centering
    \scalebox{1}{
    \begin{tabular}{|cccccc|}
      \hline {Treatment} & {Anneal [mins]} & {Vacuum [Pa]} & {O$_{2}$ [Pa]} & {Temperature [$^\circ$C]} & {[hkl]} \\
      \hline {SN1}   &  $720$  &   $1\times10^{-2}$ & $-$ & $350$ & $[111]$ \\
      \hline {SN2}   &  $10$  &   $-$ & $2\times10^{4}$ & $150$ & $[111]$ \\
      \hline {SN3}   &  $30$  &   $-$ & $2\times10^{4}$ & $150$ & $[111]$ \\
      \hline {SN4}   &  $30$  &   $-$ & $2\times10^{4}$ & $150$ & $[100]$ \\ \hline 
    \end{tabular}}
    \caption{UO$_{2+x}$ post growth thin film annealing conditions.}
    \label{tab:Table1}
\end{table}

For the XPS, the same heat treatments were replicated by placing the samples in the hot zone of a tube furnace under flowing O$_{2}$ for treatments SN$2-4$, and by annealing under the base vacuum of the XPS characterisation chamber ($< 1\times10^{-9}\:\mathrm{Pa}$) to replicate the SN$1$ treatment. While care was taken to ensure consistent parameters were used in these two annealing approaches, some variation is unavoidable and this informs the subsequent analysis. 

To establish the phase and crystal structure of the as-grown thin films, ex-situ XRD characterisation was carried out using a PANalytical X’Pert Pro MPD diffractometer with a Cu K$_{\alpha}$ tube source $\left(E=8.04\;\mathrm{keV}\right)$ with the samples mounted on a standard Eulerian cradle. For XRD characterisation in-situ to the annealing processes, an Anton Paar HTK-1200N high temperature oven was used where diffraction measurements were collected through a Kapton window.

Measurements of the U and O valence species were collected using XPS at the NanoESCA Facility (University of Bristol, UK). The U$-4f$ and O$-1s$ peaks were characterised using a monochromatic Al K$_{\alpha}$ X-ray source $\left(E=1.4867\;\mathrm{keV}\right)$, with data collected using a Scienta-Omicron Argus analyser with a pass energy of $20\;$ eV. To avoid sampling the hyperstoichiometric surface layer known to form in UO$_{2+x}$ systems \cite{Stubbs2015,Spurgeon2019} the spectra of samples annealed in O$_{2}$ were collected before and after sputtering the surface for 2 minutes with Ar ions at $500 \;\mathrm{V}$ (approximately $7\;\mu\mathrm{A}$ total sputter flux) to probe below this $\approx20$ \AA$\:$ hyperstoichiometric surface ``capping" layer. This sputtering time and acceleration voltage combination was chosen to minimise disruption to the underlying U:O stoichiometry that may arise due to preferential sputtering of O. Subsequent analysis relies on the approximation that this region is representative of the remaining bulk of the thin film, and so the calculated stoichiometry values are quoted as approximates. These stoichiometry estimates are calculated as the atomic ratio of the O-$1s$ and U-$4f$ peaks. All XPS data processing and fitting was completed using CasaXPS \cite{Fairley2021} using standard methods from contemporary studies \cite{IltonBagus2011,Ilton2017,Maslakov2018,LawrenceBright2018,Harding2023}. Prior to extracting atomic ratios, data were corrected for the transmission function of the instrument, and theoretical atomic cross-sections \cite{Scofield1976} were used for relative sensitivity factors.

\par

XAFS measurements were collected around the U $L_{3}$ absorption edge $\left(E=17.166\;\mathrm{keV}\right)$ at beam line 2-2 at the Stanford Synchrotron Radiation Lightsource, California USA, with the incident beam monochromated and calibrated using the $K$ absorption edge of an yttrium foil standard $\left(E=17.0384\;\mathrm{keV}\right)$. Samples were mounted in a vacuum sealed cylinder inside of a liquid He cryostat for temperature dependent measurements between $5 − 300\;\mathrm{K}$, with fluorescence spectra collected using a Si drift detector positioned at 45$^{\circ}$ relative to the sample normal. This detection mode is necessitated by the thin film morphology of the sample, as the millimetre thick substrate would attenuate any signal if detected in transmission mode. No beam damage to the samples was observed, and the spectra showed the expected shift in the absorption edge energy $\left(\approx 1 \;\mathrm{eV}\right)$ after warming the sample from $5 − 300\;\mathrm{K}$. Spectra were averaged over $20$ repetitions to improve the signal-to-noise ratio of the observed EXAFS oscillations within the absorption data. 

\par 

Diffraction peaks from the single crystal CaF$_{2}$ substrates were observed decorating the XAFS spectra, which if included in the energy range of the EXAFS signal processing would introduce erroneous components to the Fourier analysis. An illustrative example of such a feature is plotted in the Supplementary Information as Figure S1. To minimise the impact of these features on the results of this study, a conservative window in wavenumber $k$ was chosen of $\Delta k = 3 - 8.8$ \AA$^{-1}$. As such, the raw fluorescence spectra $\mu\left(E\right)$ measured in these experiments were normalised relative to pre-edge and post-edge intensity to provide the normalised absorption $\chi\left(E\right)$. The first derivative of the absorption edge intensity was used to define $E_{0}$, from which the wavenumber $k$ was calculated. $\chi\left(E\right)$ was then converted to $\chi\left(k\right)$, with the data within the previously detailed $\Delta k$ range Fourier transformed to provide the absorption as a function of inter-atomic distance ${R}$ from the absorbing U sites, $\chi\left(R\right)$. This normalisation and spectra processing methodology used standard procedures in the DEMETER software suite \cite{RavelNewville2005}. Multiple scattering path fits were then performed on the $\chi\left(R\right)$ within a range $\Delta R = 1-5$ \AA$\:$ using FEFF6 \cite{Zabinsky1995} to quantify inter-atomic distances, using a bulk stoichiometric UO$_{2}$ room temperature reference structure \cite{Barrett1982}.

\section{Results \& Discussion}

The results of this thin film synthesis, processing and characterisation experiment will be presented in two parts; first focussed on the as-grown and vacuum annealed thin films relative to bulk UO$_{2}$, and then on the annealed samples series. Both will feature XRD and XAFS measurements, while the XPS measurements are presented as part of the UO$_{2+x}$ study. 

\subsection{As-grown and vacuum annealed UO$_{2}$ thin films}

\begin{figure}[h!]
\centering
\begin{subfigure}{.4\textwidth}
  \centering
  \includegraphics[width=.8\linewidth]{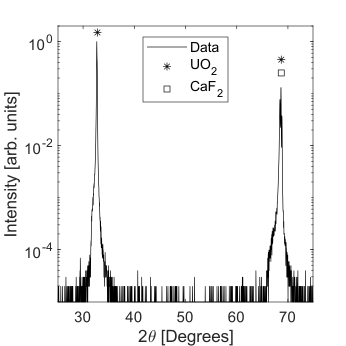}
  \caption{}
  \label{fig:Fig1_A}
\end{subfigure}%
\begin{subfigure}{.4\textwidth}
  \centering
  \includegraphics[width=.8\linewidth]{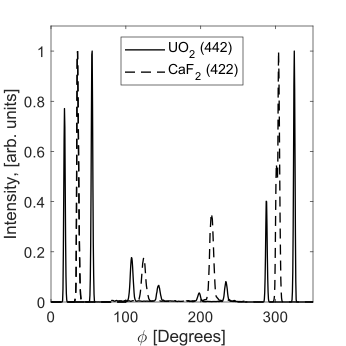}
  \caption{}
  \label{fig:Fig1_B}
\end{subfigure}
\caption{XRD of an epitaxial UO$_{2}$ thin film deposited onto a CaF$_{2}$ substrate. (a) A $\theta-2\theta$ scan, with peak positions from both the thin film and substrate labelled with symbols. (b) An off-specular $\phi$ scan, showing reflections from the thin film (solid lines) and substrate (dashed lines).}
\label{fig:Fig1}
\end{figure}

The synthesis of epitaxial UO$_{2}$ single crystals using the employed method is well established \cite{Springell2015,Rennie2018,Springell2022}, with the growth conditions employed reliably producing single crystal films with the desired crystallographic orientation. XRD data collected for one such film are shown in Figure \ref{fig:Fig1}, with Figure \ref{fig:Fig1_A} plotting the $\theta-2\theta$ scan utilised for phase identification of a $[100]$ oriented UO$_{2}$ thin film with a calculated lattice parameter of $a_{0}=5.47\pm0.01\:$\AA. Figure \ref{fig:Fig1_B} is an azimuthal $\phi$ scan over two off-specular reflections within the UO$_{2}$ thin film and CaF$_{2}$ substrate, highlighting the highly crystalline character and the epitaxial matching of the two layers.

\par 

\begin{figure}[h!]
\centering
\begin{subfigure}{.4\textwidth}
  \centering
  \includegraphics[width=.8\linewidth]{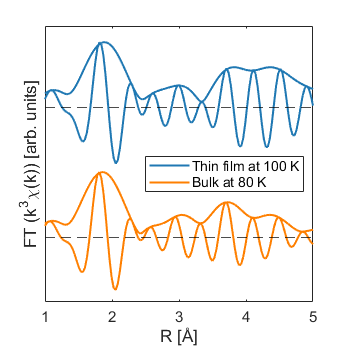}
  \caption{}
  \label{fig:Fig2_A}
\end{subfigure}%
\begin{subfigure}{.4\textwidth}
  \centering
  \includegraphics[width=.8\linewidth]{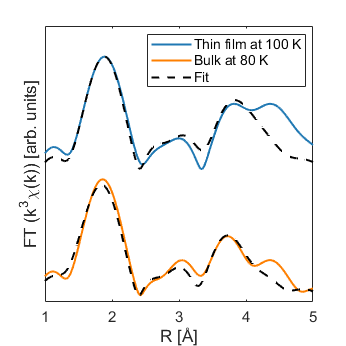}
  \caption{}
  \label{fig:Fig2_B}
\end{subfigure}
\caption{Fourier transformed U $L_{3}$ edge EXAFS $\chi\left(R\right)$ of a vacuum annealed epitaxial thin film UO$_{2}$ (blue) and bulk powder UO$_{2}$ (orange). (a) The modulus and real component of $\chi\left(R\right)$, with a dashed horizontal line representing the respective zero level. In the modulus the peaks correspond to scattering paths to nearest neighbours, while the real component of the Fourier transform highlights spectral components as a function of $R$. (b) The modulus of $\chi\left(R\right)$ shown relative to the multiple-scattering fit model (dashed lines).}
\label{fig:Fig2}
\end{figure}

EXAFS measurements of a nominally stoichiometric UO$_{2}$ thin film after vacuum annealing (blue data) relative to a bulk UO$_{2}$ reference sample (orange data) are plotted as Figure \ref{fig:Fig2}. The unfitted modulus of $\chi\left(R\right)$ for both samples is plotted along with the respective real components of $\chi\left(R\right)$ in Figure \ref{fig:Fig2_A}. Despite the $20\: \mathrm{K}$ temperature offset between these two data sets, both sample morphologies reproduce the main spectral features well. 

\par 

\par 

The plotted modulus of $\chi\left(R\right)$, which appears as an envelope to the real component of $\chi\left(R\right)$, is analogous to a partial pair-distribution function, with each peak in inter-atomic distance $R$ corresponding to scattering off of a nearest neighbour atom to the absorbing U atoms. The agreement between the two sets of data of Figure \ref{fig:Fig2_A} indicates that the local structure of the thin film and bulk powder samples are similar. The real component plot of Figure \ref{fig:Fig2_A} highlights that the signals contributing to the similar moduli of $\chi\left(R\right)$  are subtly different, particularly in the $2-3.5$ \AA\; range, suggesting that different multiple scattering processes contribute towards the observed EXAFS in the thin film relative to the bulk sample. In Figure \ref{fig:Fig2_B} the modulus of $\chi\left(R\right)$ for the thin film and bulk UO$_{2}$ samples is fitted to the expected inter-atomic distances of UO$_{2}$, showing that both data sets are well captured by the expected stoichiometric UO$_{2}$ local structure. The corresponding $\chi\left(k\right)$ representation of the data of Figure \ref{fig:Fig2} is included in the Supplementary Information as Figure S2.

\subsection{Annealed UO$_{2+x}$ thin films}

\begin{figure}[h!]
  \begin{minipage}[c]{.49\linewidth}
    \centering
    \includegraphics[width=0.8\linewidth]{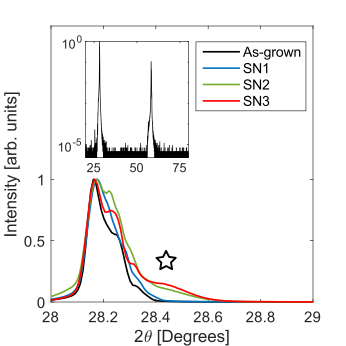}
    \captionsetup{width=.8\linewidth}
    \captionof{figure}{XRD of as-grown (black) UO$_{2}$ thin films with in-situ $\theta-2\theta$ characterisation as a function of vacuum annealing (blue) and sequentially longer oxygen annealing treatments (green, red). A non-indexed peak (star) emerged with O$_{2}$ annealing.}
    \label{fig:Fig3}
  \end{minipage}
  \hfill
  \begin{minipage}[c]{.49\linewidth}
    \centering
    \vspace{0cm}
     \scalebox{0.75}{
    \begin{tabular}{|cc|}
      \hline {Treatment} & {$a_{0}$ [\AA]} \\
      \hline {As-grown} & $5.47\pm0.01$ \\
      \hline {SN1}   &  $5.46\pm0.03$  \\
      \hline {SN2}   &  $5.47\pm0.01$  \\
      \hline {SN3}   &  $5.47\pm0.06$  \\ \hline
    \end{tabular}}
    \captionsetup{width=.8\linewidth}
    \captionof{table}{Lattice parameters calculated from XRD patterns for as-grown and annealed UO$_{2}$ thin films.}
    \label{tab:Table2}
  \end{minipage}
  \hfill
\end{figure}

As-grown UO$_{2}$ thin films were processed as detailed in Table \ref{tab:Table1} to alter the oxygen content to see what affect this has on the crystal and local structure of the system. For the XRD and XAFS studies, the same sample batch was used with XRD characterisation carried out in-situ to the annealing process, with this plotted as Figure \ref{fig:Fig3}. Here it is evident that the as-grown material again has the expected face-centred cubic structure of UO$_{2}$ (black data, inset plot), but that on magnifying the scale over the $\left(111\right)$ reflection, the impact of the annealing treatment is apparent. 

\par 

To reduce hyperstoichiometry in the as-grown thin film, the SN$1$ sample (blue data) was annealed under vacuum, resulting in the $\left(111\right)$ reflection shifting to higher $2\theta$ as the lattice contracts. Annealing in O$_{2}$ results in a minor broadening of the Bragg reflection, but no statistically significant shift in $2\theta$ relative to the as-grown material is observed. The corresponding lattice parameters are calculated from pseudo-Voigt fits to the peak centres, and are tabulated as Table \ref{tab:Table2}. Here it is evident that although the peaks in Figure \ref{fig:Fig3} become marginally broader upon annealing in O$_{2}$, there is not a shift in the cubic lattice parameter within the uncertainty of the measurement. It is notable that a secondary peak, marked by the star in Figure \ref{fig:Fig3}, grows with increasing O$_{2}$ annealing time, which is not commensurate with a distorted cubic UO$_{2+x}$ structure. 

\par

The data of Figure \ref{fig:Fig3} and the tabulated lattice parameters of Table \ref{tab:Table2} also indicate that the thin films have not undergone a phase transition, and that any  difference to the crystal structure resulting of this treatment would be subtle. The lattice parameter of UO$_{2}$ is determined by the second nearest neighbour U-U inter-atomic distance, and thus it requires a considerable addition of interstitial O to influence the lattice parameter. From this XRD characterisation, we can surmise that any hyperstoichiometry in these epitaxial thin films for the SN$2-3$ sample treatments is low, but this cannot be directly quantified by the information content of XRD.  

\par

\begin{figure}[h!]
\centering
\begin{subfigure}{.3\textwidth}
  \centering
  \includegraphics[width=1\linewidth]{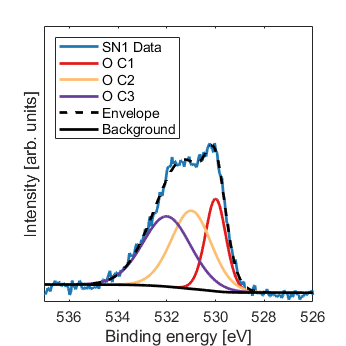}
  \caption{}
  \label{fig:Fig4A}
\end{subfigure}
\begin{subfigure}{.3\textwidth}
  \centering
  \includegraphics[width=1\linewidth]{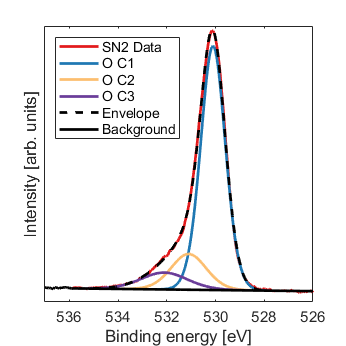}
  \caption{}
  \label{fig:Fig4B}
\end{subfigure}
\begin{subfigure}{.3\textwidth}
  \centering
  \includegraphics[width=1\linewidth]{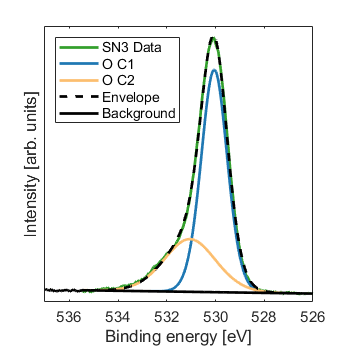}
  \caption{}
  \label{fig:Fig4C}
\end{subfigure}
\begin{subfigure}{.3\textwidth}
  \centering
  \includegraphics[width=1\linewidth]{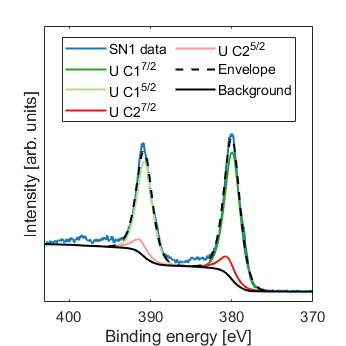}
  \caption{}
  \label{fig:Fig4D}
\end{subfigure}
\begin{subfigure}{.3\textwidth}
  \centering
  \includegraphics[width=1\linewidth]{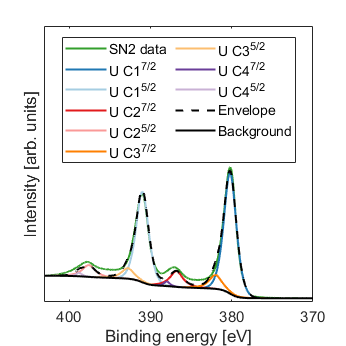}
  \caption{}
  \label{fig:Fig4E}
\end{subfigure}
\begin{subfigure}{.3\textwidth}
  \centering
  \includegraphics[width=1\linewidth]{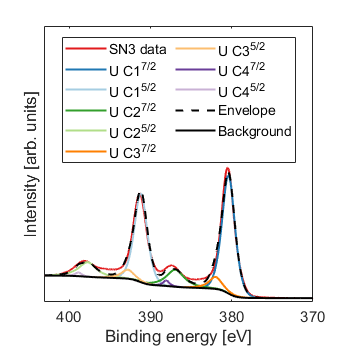}
  \caption{}
  \label{fig:Fig4F}
\end{subfigure}
\caption{X-ray photoemission spectra characterisation of annealed UO$_{2}$ epitaxial thin films. (a-c) O-$1s$ peak characterisation for SN$1-3$ respectively. (d-f) U-$4f$ peak characterisation for SN$1-3$ respectively. Each fit makes use of a Shirley background profile. Component parameters are detailed in Table \ref{tab:Table3}.}
\label{fig:Fig4}
\end{figure}

To further explore the impact of these annealing treatments on the thin film oxygen content, the surface stoichiometry was investigated by characterising the U and O valence speciation with XPS for three UO$_{2}$ $[111]$ oriented thin films processed in keeping with the samples characterised in Figure \ref{fig:Fig3}. The data for the O$-1s$ characterisation of vacuum annealed (SN$1$) and annealed in oxygen and surface sputtered with Ar$^{+}$ ions (SN$2-3$) are plotted as Figure \ref{fig:Fig4A}-\ref{fig:Fig4C}, with the corresponding U$-4f$ data plotted as Figure \ref{fig:Fig4D}-\ref{fig:Fig4F}. All fit parameters are tabulated as Table \ref{tab:Table3}. 

\par 

\par 

\begin{table}[h!]
\renewcommand{\arraystretch}{1}
    \centering
    \scalebox{0.75}{
    \begin{tabular}{|c|cccc|c|}
      \hline {Treatment} & {Component} & {Position [eV]} & {FWHM} & {\% GL} & {O:U Ratio} \\
      \hline  & {U C${1}\:^{{7/2}}$} & {$379.89$} & {$2.34$} & {$60$} & \\
                   & {U C${1}\:^{{5/2}}$} & {$390.69$} & {$2.34$} & {$60$} & \\
                   & {U C${2}\:^{{7/2}}$} & {$380.50$} & {$0.82$} & {$60$} & \\
              {SN1}      & {U C${2}\:^{{5/2}}$} & {$391.30$} & {$0.82$} & {$60$} & {$2.07\pm0.04$} \\
              \cline{2-5}     & {O C${1}$} & {$529.99$} & {$0.96$} & {$30$} & \\
                   & {O C${2}$} & {$530.98$} & {$2.16$} & {$30$} & \\
                   & {O C${3}$} & {$532.00$} & {$2.47$} & {$30$} & \\ 
      \hline  & {U C$_{1}\:^{{7/2}}$} & {$380.21$} & {$1.83$} & {$60$} & \\
                   & {U C$_{1}\:^{{5/2}}$} & {$391.01$} & {$1.83$} & {$60$} & \\
                   & {U C$_{2}\:^{{7/2}}$} & {$382.58$} & {$1.83$} & {$60$} & \\
                   & {U C$_{2}\:^{{5/2}}$} & {$392.14$} & {$1.83$} & {$60$} &  \\
              & {U C$_{3}\:^{{7/2}}$} & {$380.21$} & {$1.87$} & {$60$} & \\
               {SN2} & {U C$_{3}\:^{{5/2}}$} & {$391.01$} & {$1.87$} & {$60$} & {$2.17\pm0.02$} \\
                   & {U C$_{4}\:^{{7/2}}$} & {$382.58$} & {$0.71$} & {$60$} & \\
                    & {U C$_{4}\:^{{5/2}}$} & {$392.14$} & {$0.71$} & {$60$} & \\
              \cline{2-5}     & {O C$_{1}$} & {$530.10$} & {$1.19$} & {$30$} & \\
                   & {O C$_{2}$} & {$531.10$} & {$1.70$} & {$30$} & \\
                   & {O C$_{3}$} & {$532.10$} & {$2.27$} & {$30$} & \\ 
      \hline  & {U C$_{1}\:^{{7/2}}$} & {$380.4$} & {$2.00$} & {$60$} & \\
                   & {U C$_{1}\:^{{5/2}}$} & {$391.2$} & {$2.00$} & {$60$} & \\
                   & {U C$_{2}\:^{{7/2}}$} & {$380.48$} & {$2.00$} & {$60$} &  \\
                   & {U C$_{2}\:^{{5/2}}$} & {$391.68$} & {$2.00$} & {$60$} &  \\
             {SN3} & {U C$_{3}\:^{{7/2}}$} & {$386.60$} & {$2.57$} & {$60$} & {$2.20\pm0.02$} \\
                   & {U C$_{3}\:^{{5/2}}$} & {$397.4$} & {$2.57$} & {$60$} &  \\
            & {U C$_{4}\:^{{7/2}}$} & {$388.00$} & {$1.21$} & {$60$} &  \\
                   & {U C$_{4}\:^{{5/2}}$} & {$398.40$} & {$1.21$} & {$60$} &  \\
               \cline{2-5}  & {O C$_{1}$} & {$530.05$} & {$1.26$} & {$30$} & \\
                   & {O C$_{2}$} & {$531.05$} & {$2.45$} & {$30$} & \\ \hline
    \end{tabular}}
    \caption{Table of component parameters used in the fitting of the XPS data for the U-$4f$ and O-$1s$ peaks. Satellite peaks of the U-$4f$ C${1}$ and C${2}$ components were modelled as C${3}$ and C${4}$ respectively, with $\Delta \mathrm{BE} = 6-8\:\mathrm{eV}$. The O-$1s$ components are interpreted as lattice (C$1$), chemisorbed (C$2$), and physisorbed (C$3$) O environments respectively.}
    \label{tab:Table3}
\end{table}

\par 

\begin{figure}[h!]
\centering
\begin{subfigure}{.3\textwidth}
  \centering
  \includegraphics[width=1\linewidth]{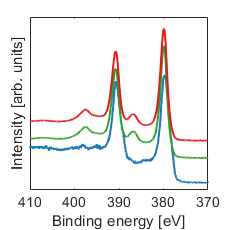}
  \caption{}
  \label{fig:Fig5A}
\end{subfigure}
\begin{subfigure}{.3\textwidth}
  \centering
  \includegraphics[width=1\linewidth]{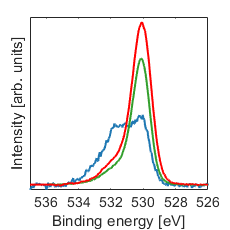}
  \caption{}
  \label{fig:Fig5B}
\end{subfigure}
\begin{subfigure}{.3\textwidth}
  \centering
  \includegraphics[width=1\linewidth]{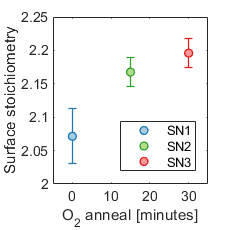}
  \caption{}
  \label{fig:Fig5C}
\end{subfigure}
\caption{X-ray photoemission spectra characterisation of annealed UO$_{2}$ epitaxial thin films. Note the SN$1$ data (blue) in (a-b) have been scaled by a factor of $50\times$ to aide comparison. (a) U-$4f$ peak characterisation. (b) O-$1s$ peak characterisation. (c) Estimated surface stoichiometry from the fitted U-$4f$ and O-$1s$ peak areas.}
\label{fig:Fig5}
\end{figure}

To aide comparison on spectral features, U-$4f$ peaks, the O-$1s$ peaks, and the estimated surface stoichiometry of the thin films are plotted as Figure \ref{fig:Fig5}. From inspection of Figure \ref{fig:Fig5A}, the SN$1$ data shows a significant attenuation of the U$-4f$ signal relative to that of SN$2-3$, with no clear evidence of the satellite peaks. This corresponds well with the SN$1$ O$-1s$ spectra in Figure \ref{fig:Fig5B}, which has a clear secondary peak component centred around $BE=532\:\mathrm{eV}$ in addition to a main component at $BE=530\:\mathrm{eV}$ attributed to the lattice O. This secondary component is attributed to physisorbed O species on the film surface, despite the vacuum conditions used for the annealing treatment. Upon annealing in O$_{2}$ and sputtering the surface, the U$-4f$ peaks shift subtly but adopt a more expected peak structure for a UO$_{2+x}$ surface, while the O$-1s$ spectral components becomes less clearly binary with the physisorbed component decreasing drastically, leaving a minor chemisorbed O shoulder at $BE=531\:\mathrm{eV}$.  next to the lattice O component by the SN$3$ treatment. Figures \ref{fig:Fig5A}-\ref{fig:Fig5B} demonstrate the responsiveness of the thin films to the oxidation treatment, while also highlighting the limitations of this surface sensitive analysis technique for such reactive oxide species. 

\par 

\par 

The surface stoichiometry of the epitaxial thin films are estimated as the ratio of the lattice O to U peak intensity, with the results of this analysis plotted as Figure \ref{fig:Fig5C}. Here it is evident that whilst SN$1$ has a lower estimated surface stoichiometry than the O$_{2}$ annealed samples, the reduced thin film retains a level of hyperstoichiometry of $x\approx0.07$. This could be the result of the previously noted surface contamination of the O-$1s$ peak influencing the detected intensity of the component assigned to the lattice oxygen, but it is difficult to be definitive for such surface sensitive measurements. The SN$2$ and SN$3$ samples show an increase in their respective surface stoichiometry estimates with annealing time, but neither is found to pass beyond $x\approx0.20$, which is below the stoichiometry commonly associated with the onset to the U$_{4}$O$_{9}$ phase of $x\geq0.25$. It is noted that this analysis relies on accurate photoelectron cross-sections, and although uncertainties in this quantity may contribute a systematic error of $\approx5\%$, the reported trend in the O:U ratio is robust.

\par 

\begin{figure}[h!]
\centering
    \begin{subfigure}{.4\textwidth}
        \centering
        \includegraphics[width=0.8\linewidth]{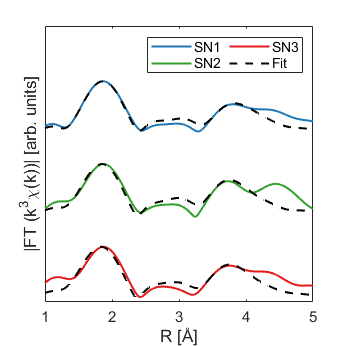}
        \caption{$5\:\mathrm{K}$}
        \label{fig:Fig6A}
    \end{subfigure}
    \begin{subfigure}{.4\textwidth}
        \centering
        \includegraphics[width=0.8\linewidth]{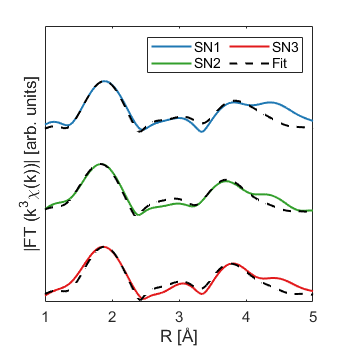}
        \caption{$100\:\mathrm{K}$}
        \label{fig:Fig6B}
    \end{subfigure}
    \par
    \begin{subfigure}{.4\textwidth}
        \centering
        \includegraphics[width=0.8\linewidth]{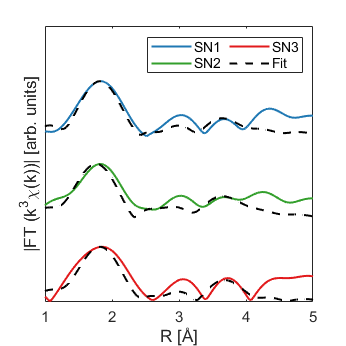}
        \caption{$200\:\mathrm{K}$}
        \label{fig:Fig6C}
    \end{subfigure}
    \begin{subfigure}{.4\textwidth}
        \centering
        \includegraphics[width=0.8\linewidth]{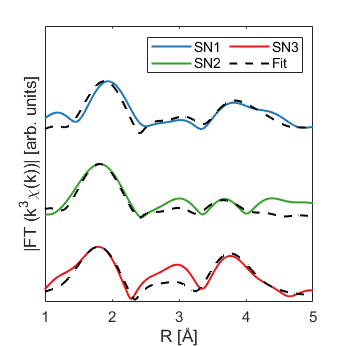}
        \caption{$300\:\mathrm{K}$}
        \label{fig:Fig6D}
    \end{subfigure}
    \caption{Magnitude of $\chi\left(R\right)$ for UO$_{2+x}$ thin film samples SN$1-3$. Each plot shows spectra collected at different sample temperatures. Multiple-scattering fits are shown as dashed black lines, with the corresponding fit parameters tabulated in the Supplementary Information as Tables S$1-3$.}
    \label{fig:Fig6}
\end{figure}

The XAFS measurements of the U $L_{3}$ absorption edge are volume averaged, and thus analysis of the EXAFS within the XAFS signal provides a probe of the local structure that is not surface limited, with an inter-atomic distance sensitivity range of $\mathrm{R}\leq5.5\:$\AA. The magnitude of $\chi\left(R\right)$ for the three previously discussed sample treatments (SN$1-3$) are plotted as Figure \ref{fig:Fig6}, with each plot representing spectra collected at different sample temperatures. The fits to the data (dashed black lines) are limited to the U-O$_{1}$, U-U$_{1}$ and U-O$_{2}$ nearest neighbour scattering paths for pristine bulk UO$_{2}$, and thus spectral weight unaccounted for by these fits is suggestive of scattering from non-stoichiometric atomic sites. The corresponding $\chi\left(k\right)$ representation of each of these data are included in the Supplementary Information as Figure S3.

The plots of Figure \ref{fig:Fig6} highlight that at $5\:$K, the EXAFS shows a good agreement between the stoichiometric UO$_{2}$ scattering model for all samples between $1-5\:$\AA. As the sample temperature is increased, the agreement between the model and the data generally decreases, with the most disagreement seen in the SN$2$ data above $200\:$K. Notably, the SN$3$ data is particularly poorly fitted in the first peak, corresponding to scattering from the U-O$_{1}$ path, at $200\:$K (red data, Figure \ref{fig:Fig6C}).

\par 

\begin{figure}[h!]
\centering
    \begin{subfigure}{.4\textwidth}
        \centering
        \includegraphics[width=0.8\linewidth]{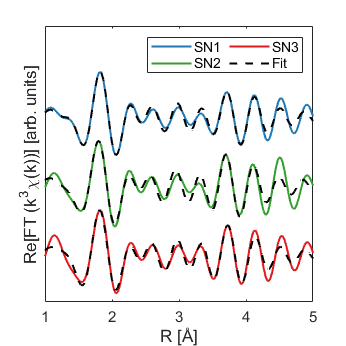}
        \caption{$5\:\mathrm{K}$}
        \label{fig:Fig7A}
    \end{subfigure}
    \begin{subfigure}{.4\textwidth}
        \centering
        \includegraphics[width=0.8\linewidth]{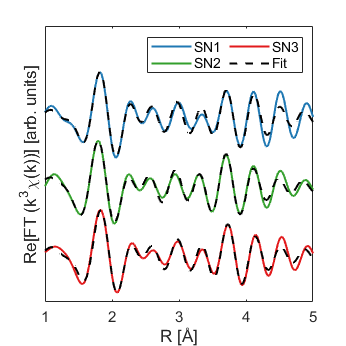}
        \caption{$100\:\mathrm{K}$}
        \label{fig:Fig7B}
    \end{subfigure}
    \par
    \begin{subfigure}{.4\textwidth}
        \centering
        \includegraphics[width=0.8\linewidth]{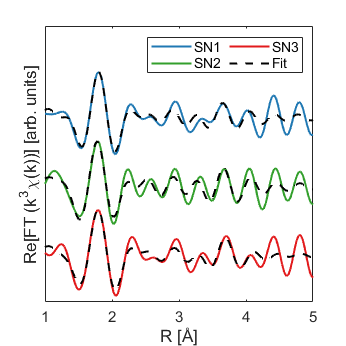}
        \caption{$200\:\mathrm{K}$}
        \label{fig:Fig7C}
    \end{subfigure}
    \begin{subfigure}{.4\textwidth}
        \centering
        \includegraphics[width=0.8\linewidth]{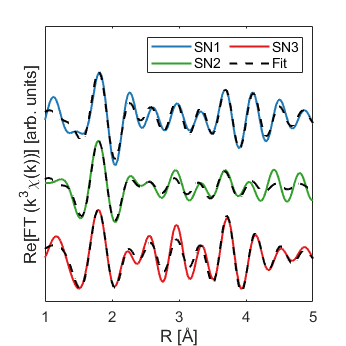}
        \caption{$300\:\mathrm{K}$}
        \label{fig:Fig7D}
    \end{subfigure}
    \caption{Real component of $\chi\left(R\right)$ for UO$_{2+x}$ thin film samples SN$1-3$. Each plot shows spectra collected at different sample temperatures. Multiple-scattering fits to each of the spectra are included as dashed black lines.}
    \label{fig:Fig7}
\end{figure}

The real component of $\chi\left(R\right)$, and the corresponding multiple scattering fits, are plotted for SN$1-3$ as Figure \ref{fig:Fig7}. In keeping with the trends observed in Figure \ref{fig:Fig6}, the agreement between the UO$_{2}$ multiple scattering model is higher at low temperature, with this generally decreasing as the temperature of the sample increases. Notably here the SN$3$ $300\:$K data (red data, Figure \ref{fig:Fig7D}) shows a good agreement in the real component of $\chi\left(R\right)$, with SN$2$ showing a greater disagreement across the $200-300\:$K range. Importantly, the plots of Figure \ref{fig:Fig7} highlight that each EXAFS measurement has distinct features that are not common to every sample, and that these evolve with the measurement temperature. This indicates that there is some reordering of the scattering sites within the material as the sample temperature changes, with different configurations being more or less well captured by the stoichiometric UO$_{2}$ model.  

\par 

\begin{figure}[h!]
    \centering
    \begin{subfigure}{.49\textwidth}
        \centering
        \includegraphics[width=0.8\linewidth]{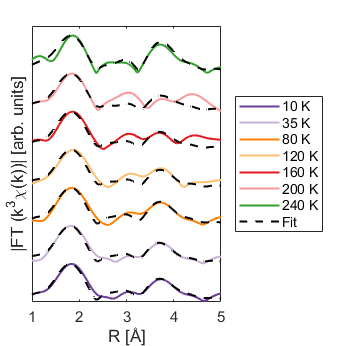}
    \caption{}
    \label{fig:Fig8A}
    \end{subfigure}
    \begin{subfigure}{.49\textwidth}
        \centering
        \includegraphics[width=0.8\linewidth]{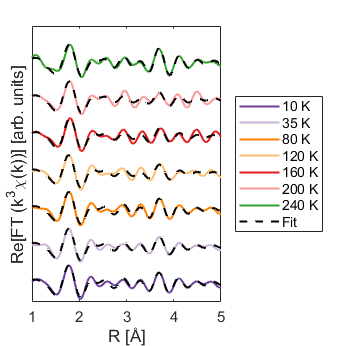}
    \caption{}
    \label{fig:Fig8B}
    \end{subfigure}
    \caption{Absorption spectra for the UO$_{2+x}$ thin film samples SN$4$, plotted as a function of sample temperature.  The multiple-scattering fits to each of the spectra is included as dashed black lines. (a) Magnitude and (b) Real component, of the Fourier transformed absorption spectra, respectively.}
    \label{fig:Fig8}
\end{figure}

To further probe the temperature dependence of the EXAFS, a second sample was processed using the same treatment of SN$3$, grown on $[100]$ oriented CaF$_{2}$ to see if the variation in substrate orientation influences the observed temperature dependence in the local structure. For clarity this sample is termed SN$4$, but experienced the same sample treatment as SN$3$ as previously detailed. These measurements were collected at finer temperature increments to better probe temperature dependent spectral evolution, but were otherwise measured using the same method as the spectra plotted in Figure \ref{fig:Fig7} and Figure \ref{fig:Fig8}. The SN$4$ spectra are fitted with the same stoichiometric UO$_{2}$ model used for the SN$1-3$ analysis, to determine the deviation from the ideal UO$_{2}$ structure without allowing for additional scattering paths within the unit cell. The corresponding $\chi\left(R\right)$ measurements are plotted as Figure \ref{fig:Fig8}, with $\chi\left(k\right)$ plotted in the Supplementary Information as Figure S4.

\par 

Both the magnitude and real component of $\chi\left(R\right)$ plotted as Figures \ref{fig:Fig8A} and \ref{fig:Fig8B} respectively show a similar evolution with sample temperature as observed in the four sets of data plotted for SN$3$ in Figures \ref{fig:Fig6}-\ref{fig:Fig7}. In both of the representations of the data in Figure \ref{fig:Fig8} there is a closer agreement between the UO$_{2}$ model and the data at low temperature and near room temperature, with the disagreement more pronounced in the $80-200\:$K range. While this appears more dramatic in the magnitude plot of Figure \ref{fig:Fig8A}, the real component representation of Figure \ref{fig:Fig8B} demonstrates that this arises due to new components in the Fourier transform emerging in this range of sample temperatures, which cannot be accounted for in this three scattering path UO$_{2}$ basis. 

\par 

\begin{figure}[h!]
\centering
    \begin{subfigure}{.45\textwidth}
        \centering
        \includegraphics[width=0.8\linewidth]{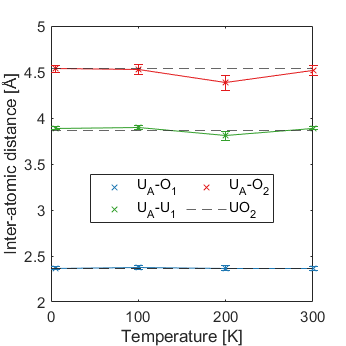}
        \caption{SN1}
        \label{fig:Fig9A}
    \end{subfigure}
    \begin{subfigure}{.45\textwidth}
        \centering
        \includegraphics[width=0.8\linewidth]{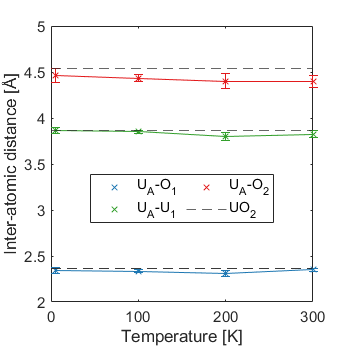}
        \caption{SN2}
        \label{fig:Fig9B}
    \end{subfigure}
    \par
    \begin{subfigure}{.45\textwidth}
        \centering
        \includegraphics[width=0.8\linewidth]{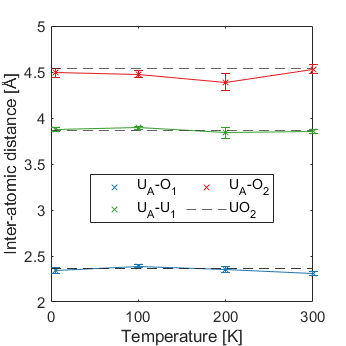}
        \caption{SN3}
        \label{fig:Fig9C}
    \end{subfigure}
    \begin{subfigure}{.45\textwidth}
        \centering
        \includegraphics[width=0.8\linewidth]{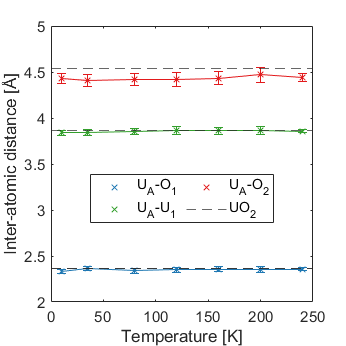}
        \caption{SN4}
        \label{fig:Fig9D}
    \end{subfigure}
    \caption{Fitted inter-atomic distances for the UO$_{2+x}$ EXAFS. Here each modelled scattering path is shown relative to the room temperature value for bulk UO$_{2}$, with the nearest neighbour U-O (blue) and U-U (green) paths and the second nearest neighbour U-O (red) path shown for various sample temperatures.}
    \label{fig:Fig9}
\end{figure}

The quantitative output of the multiple scattering fits are tabulated in full in the Supplementary Information as Tables S$1-4$, but the fitted inter-atomic distances of the fits to samples SN$1-4$ for the three modelled scattering paths are plotted separately as Figure \ref{fig:Fig9}. Here the uncertainty on each data point is propagated from the FEFF6 fitting model, and the dashed horizontal lines correspond to the expected inter-atomic distance for that scattering path in bulk UO$_{2}$ at $300\:$K. Overall, the two O scattering paths are the most variable of the three included in these fits, with the U-O$_{2}$ path being notably more sensitive to both the O$_{2}$ annealing and the variation in the sample temperature during the measurement. This is markedly different from the nearest neighbour U-U$_{1}$ scattering path, which is largely invariant from the expected stoichiometric inter-atomic distance for each of the fits, in keeping with the findings of the XRD analysis of these samples. 

The inter-atomic distances fitted for SN$1$ agree with the bulk reference structure well, with the U-O$_{2}$ distance at $200\:$ K being a notable variation from this relative invariance with sample temperature during the measurement. The inter-atomic distances for SN$2-4$ vary more notably from the bulk room temperature structure, with this most clearly manifesting in the U-O$_{2}$ distances, but also in the U-U and U-O$_{1}$ of SN$2$ and SN$4$. It is notable that the SN$2$ U-O$_{1}$ distance varies more from the stoichiometric value than the SN$4$ values do, indicating that the degree of distortion from the stoichiometric structure does not increase monotonically with increased oxidation. 

\par 

Together, Figures \ref{fig:Fig6}-\ref{fig:Fig9} show the complexity of these mildly oxidised epitaxial thin film systems as viewed from the bulk averaged probe of XAFS, with the EXAFS analysis indicating that while the low temperature local structure appears closely aligned to that of stoichiometric UO$_{2}$, the prominence of interstitial O sites becomes more apparent with elevated temperature. Furthermore, the U-U$_{1}$ behaviour corroborates the XRD analysis well, indicating the U sublattice behaves in a similar way in these epitaxial systems as is observed in the bulk material. Finally, the sensitivity of the local structure to sample temperature is notable, and hints at a dynamic quality to the O sublattice which is most readily observed after oxidative annealing. 

\section{Conclusions}

This synthesis route for epitaxial uranium oxide systems has successfully produced phase pure UO$_{2}$, which were then tailored to differing O content through annealing treatments. While the absolute degree of stoichiometry is non-trivial to determine using non-destructive methods, this range of annealing treatments is estimated through XPS analysis to result in a stoichiometric range of $0.07 \leq x \leq 0.20$, and thus are not expected to show the U$_{4}$O$_{9}$ phase where $x\geq 0.25$. These samples instead occupy a transitional region of the UO$_{2+x}$ phase diagram, where O clustering and interstitial O site occupancy are thought to drive the observed alterations to the material properties in the bulk. The variation in the local structure with both oxidation treatment and sample temperature during measurement as probed through the EXAFS analysis highlights the efficacy of this epitaxial thin film synthesis route in replicating the complex material character of the bulk material, and suggests that such epitaxial systems could be employed as proxies to the bulk in more convoluted material studies in which the use of bulk or powdered uranium oxide materials are prohibitive. Furthermore, the observed variation in the local structure with sample temperature is suggestive of a pronounced dynamic character to the interstitial sites within the UO$_{2+x}$ O sublattice, which has been suggested in other contemporary studies of the bulk material and powdered systems. 

\section*{Acknowledgements}

JCL was funded by the UKRI Engineering and Physical Sciences Research Council (EPSRC), through the Centre for Doctoral Training in Condensed Matter Physics (CDT-CMP) grant no. EP/L015544/1, with additional funding for this work from the Transformative Science and Engineering for Nuclear Decommissioning (TRANSCEND) consortium, grant no. EP/L014041/1. This work was further bbed by the EPSRC NNUF FaRMS Facility, grant no. EP/V035495/1. Use of the Stanford Synchrotron Radiation Lightsource, SLAC National Accelerator Laboratory, is supported by the U.S. Department of Energy, Office of Science, Office of Basic Energy Sciences under contract no. DE-AC02-76SF00515. The authors also acknowledge use of the University of Bristol NanoESCA Laboratory.

\section*{Contributions}

JCL (Conceptualization, Formal analysis, Investigation, Methodology, Writing – original draft, Writing - review \& editing), SDC (Formal analysis, Investigation, Writing - review \& editing), JW (Investigation, Writing - review \& editing), LMH (Investigation, Writing - review \& editing), RN (Investigation, Writing - review \& editing), JL (Investigation, Writing - review \& editing), CB (Conceptualization, Funding acquisition, Supervision, Writing – original draft, Writing – review \& editing), RSS (Conceptualization, Funding acquisition, Supervision, Writing – review \& editing). 

\bibliographystyle{ieeetr}
\bibliography{bibtex.bib}

\newpage

\section*{Supplemental Information}
\setcounter{section}{0}
\renewcommand{\thesection}{S\arabic{section}}
\setcounter{figure}{0}
\renewcommand{\thefigure}{S\arabic{figure}}
\setcounter{table}{0}
\renewcommand{\thetable}{S\arabic{table}}
\section{Additional EXAFS Figures}

\vspace{-0.5cm}
\begin{figure}[h!]
    \centering
    \includegraphics[width=0.5\linewidth]{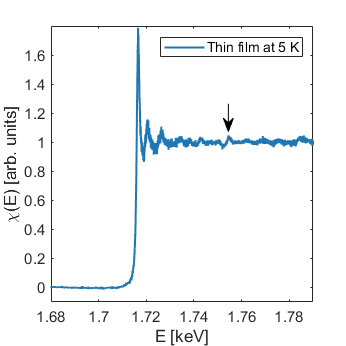}
    \caption{Normalised U L$_{3}$ edge absorption spectra for SN$1$ at $5\:$K. Highlighted with a black arrow is a diffraction peak.}
    \label{fig:SI-Fig1}
\end{figure}
\vspace{-0.5cm}
\begin{figure}[h!]
    \centering
    \includegraphics[width=0.5\linewidth]{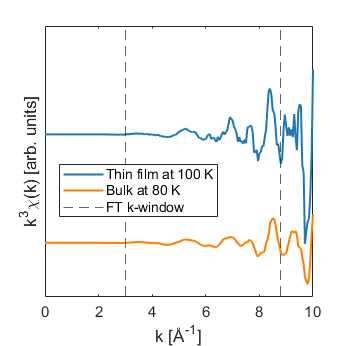}
    \caption{$\chi\left(k\right)$ for SN$1$ (blue, $100\:\mathrm{K}$) and bulk UO$_{2}$ (orange, $80\:\mathrm{K}$). Dashed vertical lines show the $\Delta k$ Fourier transform window.}
    \label{fig:SI-Fig2}
\end{figure}

\begin{figure}[h!]
    \centering
    \begin{subfigure}{.4\textwidth}
        \includegraphics[width=\textwidth]{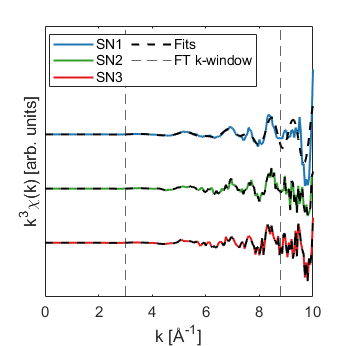}
        \caption{$5\:\mathrm{K}$}
    \end{subfigure}
    \begin{subfigure}{.4\textwidth}
        \includegraphics[width=\textwidth]{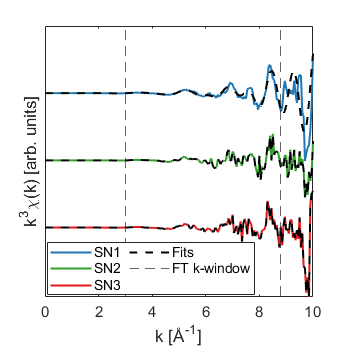}
        \caption{$100\:\mathrm{K}$}
    \end{subfigure}
    \begin{subfigure}{.4\textwidth}
        \includegraphics[width=\textwidth]{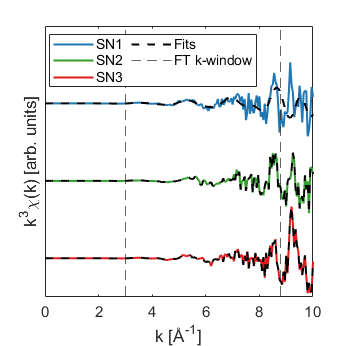}
        \caption{$200\:\mathrm{K}$}
    \end{subfigure}
    \begin{subfigure}{.4\textwidth}
        \includegraphics[width=\textwidth]{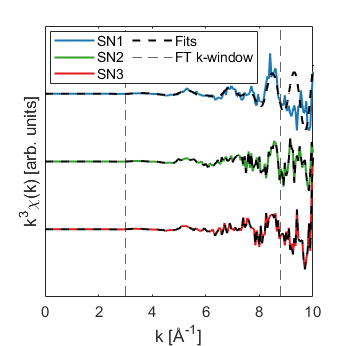}
        \caption{$300\:\mathrm{K}$}
    \end{subfigure}
    \caption{$\chi\left(k\right)$ absorption data plotted for samples SN$1-3$ for temperatures of $5, 100, 200,\: \mathrm{and} \: 300 \:\mathrm{K}$. Each spectra is plotted scaled by $k^{3}$, magnifying the high-$k$ oscillations. Included in each plot is a representation of the same $\mathrm{R}$-space multiple-scattering path fit from the $\chi\left(\mathrm{R}\right)$ data. The $\Delta k$ Fourier transform window range is denoted by the dashed vertical lines.}
    \label{fig:SI-Fig3}
\end{figure}

\vfill

\begin{figure}[h!]
    \centering
    \includegraphics[width=0.5\linewidth]{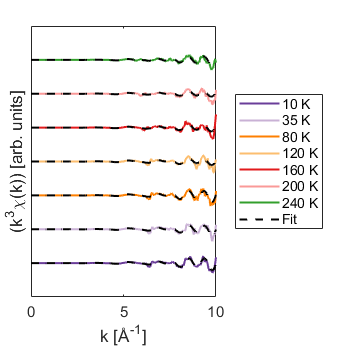}
    \caption{$\chi\left(k\right)$ absorption data plotted for sample SN$4$ for temperatures of $10-240 \:\mathrm{K}$. Each spectra is plotted scaled by $k^{3}$, to magnify the high-$k$ oscillations. Included in each plot is a representation of the same $\mathrm{R}$-space multiple-scattering path fit from the $\chi\left(\mathrm{R}\right)$ data. }
    \label{fig:SI-Fig4}
\end{figure}
\clearpage
\section{EXAFS Fitting Parameters}

\begin{table}[h!]
    \renewcommand{\arraystretch}{1.25}
    \centering
    \scalebox{0.75}{
    \begin{tabular}{|ccccc|c|c|} \hline
       {Temperature [K]} & {Path} & {${\Delta\mathrm{E_{0}}}$ [eV]} & $\sigma^{2} \mathrm{\: [\AA^{2}]}$ & {$\mathrm{R \: [\AA]}$} & Fitting $\Delta\mathrm{R \: [\AA]}$ & FT $\Delta\mathrm{k \: [\AA^{-1}]}$ \\ \hline
        & U-O$_{\mathrm{1}}$ & \multirow{3}{2.24em}{$4.58$} & {$0.0051\pm0.0012$} & {$2.366\pm0.017$} & \multirow{12}{4em}{$1.0-5.0$}  &  \multirow{12}{4em}{$3-8.8$}\\ 
       $5$ & U-U$_{\mathrm{1}}$ &  & {$0.0019\pm0.0018$} & {$3.891\pm0.016$} &  & \\ 
        & U-O$_{\mathrm{2}}$ &  & {$0.0051\pm0.0012$} & {$4.534\pm0.039$} &  & \\ \cline{1-5}
        & U-O$_{\mathrm{1}}$ &  \multirow{3}{2.24em}{$6.32$} & {$0.0056\pm0.0017$} & {$2.374\pm0.024$} &  & \\ 
       $100$ & U-U$_{\mathrm{1}}$ &  & {$0.0010\pm0.0009$} & {$3.894\pm0.021$} &  & \\ 
        & U-O$_{\mathrm{2}}$ &  & {$0.0056\pm0.0017$} & {$4.524\pm0.055$} &  & \\ \cline{1-5}
        & U-O$_{\mathrm{1}}$ &  \multirow{3}{2.24em}{$5.38$} & {$0.0061\pm0.0023$} & {$2.367\pm0.027$} &  & \\ 
       $200$ & U-U$_{\mathrm{1}}$ &  & {$0.0058\pm0.0031$} & {$3.805\pm0.053$} &  & \\ 
        & U-O$_{\mathrm{2}}$ &  & {$0.0061\pm0.0023$} & {$4.381\pm0.084$} &  & \\ \cline{1-5}
        & U-O$_{\mathrm{1}}$ &  \multirow{3}{2.24em}{$4.58$} & {$0.0068\pm0.0018$} & {$2.368\pm0.023$} &  & \\ 
       $300$ & U-U$_{\mathrm{1}}$ &  & {$0.0015\pm0.0010$} & {$3.891\pm0.020$} &  & \\ 
        & U-O$_{\mathrm{2}}$ &  & {$0.0068\pm0.0018$} & {$4.518\pm0.055$} &  & \\ \hline
    \end{tabular}}
    \caption{$\chi\left(R\right)$ fit parameters for sample SN$1$.}
    \label{tab:SI_Table1_SN1_EXAFS}
\end{table}

\begin{table}[h!]
    \renewcommand{\arraystretch}{1.25}
    \centering
    \scalebox{0.75}{
    \begin{tabular}{|ccccc|c|c|} \hline
       {Temperature [K]} & {Path} & {${\Delta\mathrm{E_{0}}}$ [eV]} & $\sigma^{2} \mathrm{\: [\AA^{2}]}$ & {$\mathrm{R \: [\AA]}$} & Fitting $\Delta\mathrm{R \: [\AA]}$ & FT $\Delta\mathrm{k \: [\AA^{-1}]}$ \\ \hline
        & U-O$_{\mathrm{1}}$ & \multirow{3}{2.24em}{$4.95$} & {$0.0084\pm0.0024$} & {$2.346\pm0.033$} & \multirow{12}{4em}{$1.0-5.0$}  &  \multirow{12}{4em}{$3-8.8$}\\ 
       $5$ & U-U$_{\mathrm{1}}$ &  & {$0.0019\pm0.0018$} & {$3.864\pm0.028$} &  & \\ 
        & U-O$_{\mathrm{2}}$ &  & {$0.0084\pm0.0024$} & {$4.462\pm0.075$} &  & \\ \cline{1-5}
        & U-O$_{\mathrm{1}}$ & \multirow{3}{2.24em}{$3.29$} & {$0.0094\pm0.0012$} & {$2.333\pm0.016$} &  & \\ 
       $100$ & U-U$_{\mathrm{1}}$ &  & {$0.0029\pm0.0009$} & {$3.848\pm0.014$} &  & \\ 
        & U-O$_{\mathrm{2}}$ &  & {$0.0094\pm0.0012$} & {$4.433\pm0.036$} &  & \\ \cline{1-5}
        & U-O$_{\mathrm{1}}$ & \multirow{3}{2.24em}{$0.84$} & {$0.0097\pm0.0022$} & {$2.309\pm0.035$} &  & \\ 
       $200$ & U-U$_{\mathrm{1}}$ &  & {$0.0073\pm0.0029$} & {$3.801\pm0.041$} &  & \\ 
        & U-O$_{\mathrm{2}}$ &  & {$0.0097\pm0.0022$} & {$4.402\pm0.081$} &  & \\ \cline{1-5}
        & U-O$_{\mathrm{1}}$ & \multirow{3}{2.24em}{$4.57$} & {$0.0069\pm0.0017$} & {$2.349\pm0.020$} &  & \\ 
       $300$ & U-U$_{\mathrm{1}}$ &  & {$0.0064\pm0.0024$} & {$3.826\pm0.038$} &  & \\ 
        & U-O$_{\mathrm{2}}$ &  & {$0.0069\pm0.0017$} & {$4.395\pm0.062$} &  & \\ \hline
    \end{tabular}}
    \caption{$\chi\left(R\right)$ fit parameters for sample SN$2$.}
    \label{tab:SI_Table1_SN2_EXAFS}
\end{table}

\begin{table}[h!]
    \renewcommand{\arraystretch}{1.25}
    \centering
    \scalebox{0.75}{
    \begin{tabular}{|ccccc|c|c|} \hline
       {Temperature [K]} & {Path} & {${\Delta\mathrm{E_{0}}}$ [eV]} & $\sigma^{2} \mathrm{\: [\AA^{2}]}$ & {$\mathrm{R \: [\AA]}$} & Fitting $\Delta\mathrm{R \: [\AA]}$ & FT $\Delta\mathrm{k \: [\AA^{-1}]}$ \\ \hline
        & U-O$_{\mathrm{1}}$ & \multirow{3}{2.24em}{$1.92$} & {$0.0082\pm0.0016$} & {$2.344\pm0.026$} & \multirow{12}{4em}{$1.0-5.0$}  &  \multirow{12}{4em}{$3-8.8$}\\ 
       $5$ & U-U$_{\mathrm{1}}$ &  & {$0.0028\pm0.0017$} & {$3.872\pm0.021$} &  & \\ 
        & U-O$_{\mathrm{2}}$ &  & {$0.0082\pm0.0016$} & {$4.490\pm0.053$} &  & \\ \cline{1-5}
        & U-O$_{\mathrm{1}}$ & \multirow{3}{2.24em}{$6.03$} & {$0.0084\pm0.0013$} & {$2.390\pm0.015$} &  & \\ 
       $100$ & U-U$_{\mathrm{1}}$ &  & {$0.0022\pm0.0010$} & {$3.895\pm0.015$} &  & \\ 
        & U-O$_{\mathrm{2}}$ &  & {$0.0084\pm0.0013$} & {$4.474\pm0.037$} &  & \\ \cline{1-5}
        & U-O$_{\mathrm{1}}$ & \multirow{3}{2.24em}{$5.45$} & {$0.0120\pm0.0026$} & {$2.356\pm0.030$} &  & \\ 
       $200$ & U-U$_{\mathrm{1}}$ &  & {$0.0109\pm0.0052$} & {$3.839\pm0.059$} &  & \\ 
        & U-O$_{\mathrm{2}}$ &  & {$0.0120\pm0.0026$} & {$4.388\pm0.093$} &  & \\ \cline{1-5}
        & U-O$_{\mathrm{1}}$ & \multirow{3}{2.24em}{$0.17$} & {$0.0108\pm0.0017$} & {$2.314\pm0.021$} &  & \\ 
       $300$ & U-U$_{\mathrm{1}}$ &  & {$0.0049\pm0.0021$} & {$3.851\pm0.021$} &  & \\ 
        & U-O$_{\mathrm{2}}$ &  & {$0.0108\pm0.0017$} & {$4.532\pm0.046$} &  & \\ \hline
    \end{tabular}}
    \caption{$\chi\left(R\right)$ fit parameters for sample SN$3$.}
    \label{tab:SI_Table1_SN3_EXAFS}
\end{table}

\begin{table}[h!]
    \renewcommand{\arraystretch}{1.25}
    \centering
    \scalebox{0.75}{
    \begin{tabular}{|ccccc|c|c|} \hline
       {Temperature [K]} & {Path} & {${\Delta\mathrm{E_{0}}}$ [eV]} & $\sigma^{2} \mathrm{\: [\AA^{2}]}$ & {$\mathrm{R \: [\AA]}$} & Fitting $\Delta\mathrm{R \: [\AA]}$ & FT $\Delta\mathrm{k \: [\AA^{-1}]}$ \\ \hline
        & U-O$_{\mathrm{1}}$ & \multirow{3}{2.24em}{$3.93$} & {$0.0126\pm0.0018$} & {$2.330\pm0.024$} &  \multirow{21}{4em}{$1.0-5.0$}  &  \multirow{21}{4em}{$3.0-8.8$} \\ 
       $10$ & U-U$_{\mathrm{1}}$ &  & {$0.0064\pm0.0017$} & {$3.841\pm0.026$} &  & \\ 
        & U-O$_{\mathrm{2}}$ &  & {$0.0126\pm0.0018$} & {$4.427\pm0.057$} &  &  \\ \cline{1-5}
        & U-O$_{\mathrm{1}}$ & \multirow{3}{2.24em}{$4.74$} & {$0.0123\pm0.0021$} & {$2.336\pm0.024$} &  &  \\ 
       $35$ & U-U$_{\mathrm{1}}$ &  & {$0.0075\pm0.0025$} & {$3.845\pm0.033$} &  &  \\ 
        & U-O$_{\mathrm{2}}$ &  & {$0.0123\pm0.0021$} & {$4.405\pm0.065$} &  &  \\ \cline{1-5}
        & U-O$_{\mathrm{1}}$ & \multirow{3}{2.24em}{$4.95$} & {$0.0126\pm0.0022$} & {$2.339\pm0.024$} &  &  \\ 
       $80$ & U-U$_{\mathrm{1}}$ &  & {$0.0064\pm0.0022$} & {$3.852\pm0.029$} &  &  \\ 
        & U-O$_{\mathrm{2}}$ &  & {$0.0126\pm0.0022$} & {$4.419\pm0.063$} &  &  \\ \cline{1-5}
        & U-O$_{\mathrm{1}}$ & \multirow{3}{2.24em}{$6.44$} & {$0.0129\pm0.0024$} & {$2.348\pm0.026$} &  &  \\ 
       $120$ & U-U$_{\mathrm{1}}$ &  & {$0.0097\pm0.0038$} & {$3.864\pm0.044$} &  &  \\ 
        & U-O$_{\mathrm{2}}$ &  & {$0.0129\pm0.0024$} & {$4.414\pm0.077$} &  &  \\ \cline{1-5}
        & U-O$_{\mathrm{1}}$ & \multirow{3}{2.24em}{$4.06$} & {$0.0133\pm0.0023$} & {$2.355\pm0.028$} &  &  \\ 
       $160$ & U-U$_{\mathrm{1}}$ &  & {$0.0092\pm0.0029$} & {$3.861\pm0.041$} &  &  \\ 
        & U-O$_{\mathrm{2}}$ &  & {$0.0133\pm0.0023$} & {$4.434\pm0.074$} &  &  \\ \cline{1-5}
        & U-O$_{\mathrm{1}}$ & \multirow{3}{2.24em}{$5.11$} & {$0.0152\pm0.0025$} & {$2.357\pm0.034$} &  &  \\ 
       $200$ & U-U$_{\mathrm{1}}$ &  & {$0.0112\pm0.0032$} & {$3.867\pm0.046$} &  &  \\ 
        & U-O$_{\mathrm{2}}$ &  & {$0.0152\pm0.0025$} & {$4.470\pm0.084$} &  &  \\ \cline{1-5}
        & U-O$_{\mathrm{1}}$ & \multirow{3}{2.24em}{$4.84$} & {$0.0189\pm0.0016$} & {$2.358\pm0.017$} &  &  \\ 
       $240$ & U-U$_{\mathrm{1}}$ &  & {$0.0086\pm0.0013$} & {$3.858\pm0.018$} &  &  \\ 
        & U-O$_{\mathrm{2}}$ &  & {$0.0189\pm0.0016$} & {$4.44\pm0.042$} & &  \\ \hline
    \end{tabular}}
    \caption{$\chi\left(R\right)$ fit parameters for sample SN$4$.}
    \label{tab:SI_Table1_SN4_EXAFS}
\end{table}

\end{document}